\documentclass[%
 reprint,
amsmath,amssymb,
prl,
]{revtex4-2}

\usepackage{graphicx}% Include figure files
\usepackage{dcolumn}% Align table columns on decimal point
\usepackage{bm}% bold math
\usepackage{float}
\usepackage[hidelinks,colorlinks=true,linkcolor=blue,citecolor=blue,allcolors=blue]{hyperref}% add hypertext capabilities
\makeatletter
\pretocmd{\NAT@open}{\begingroup\color{\@citecolor}}{}{}
\apptocmd{\NAT@close}{\endgroup}{}{}
\makeatother
\usepackage{adjustbox}

\usepackage{titlesec} 

\newcommand{\addperiod}[1]{#1\ --- \hspace{1.0em}}

\titleformat{\section}[runin]
  {\normalfont\normalsize\itshape}{}{0.5em}{\addperiod}
\titleformat{\subsection}[runin]
  {\normalfont\normalsize\itshape}{}{0.5em}{\addperiod}
\titleformat{\subsubsection}[runin]
  {\normalfont\normalsize\itshape}{}{0.5em}{\addperiod}

\titlespacing\section{0pt}{0pt plus 4pt minus 2pt}{0pt plus 2pt minus 2pt}
\titlespacing\subsection{0pt}{12pt plus 4pt minus 2pt}{0pt plus 2pt minus 2pt}
\titlespacing\subsubsection{0pt}{12pt plus 4pt minus 2pt}{0pt plus 2pt minus 2pt}

\begin{document}

%\preprint{APS/123-QED}

\title{Schrödinger's Cheshire Cat: A tabletop experiment to measure the Diósi-Penrose collapse time and demonstrate Objective Reduction (OR)}% Force line breaks with \\

\author{James Tagg}
\author{William Reid}%
\author{Daniel Carlin}
%\email{james.tagg@valiscorp.com}
\affiliation{Valis Corporation, Encinitas, California, 92024, james.tagg@valiscorp.com}%


\date{\today}% It is always \today, today,

\begin{abstract}
For nearly 100 years, the paradox of Schrödinger's Cat has remained unresolved. Why does the world we live in appear classical despite being composed of quantum particles governed by the Schrödinger wave equation? Lajos Diósi and Roger Penrose propose the wavefunction collapses because it describes two incompatible spacetimes, demonstrating an inconsistency between quantum mechanics and general relativity. To avoid this paradox, collapse must occur within Heisenberg’s time-energy uncertainty limit. Subatomic particles with low mass and correspondingly low energy collapse in years, while superposed cats would collapse almost instantaneously. We propose a table-top experiment to put two small mirrors into superposition and observe them collapse in a time consistent with the Diósi-Penrose model. We employ two techniques to perform this experiment in ambient laboratory conditions. Most experiments separate a small mass by a large distance. In contrast, we displace a large mass by a small distance where the self-energy follows an inverse square law with correspondingly high collapse times. We further use a symmetrical apparatus, where a break in symmetry indicates collapse independent of decoherence. 
\end{abstract}

\maketitle

\section{Introduction.\label{sec:1}}

Erwin Schrödinger famously suggested quantum mechanics was flawed because it would be possible to put a cat into superposition \cite{1}. In the original paper, a cat is placed in a steel box with a diabolical device: a Geiger counter arranged to break a flask of poison if it detects the decay of a small radioactive sample. The sample decays with a 50\% probability over the course of an hour. At the end of the hour, the system's wavefunction should describe a mix of a living and a dead cat. It seemed ridiculous to Schrödinger that one could scale up quantum superposition to large objects such as cats, and thus the paradox was born.

There are many ways to handle his paradox. Models, such as the Copenhagen interpretation, say the act of observing the cat causes state collapse, but this creates its own set of paradoxes \cite{2}. The Everett many-worlds interpretation \cite{3} argues that the wavefunction never collapses. In the world of my conscious experience, the cat lives, but in another, it was not so lucky.

Another way to resolve Schrödinger's paradox is to follow his original intent. The paradox is a clue to a flaw in our current understanding of quantum mechanics. Theories of this type are labeled ‘objective' because no observer is required and they should be experimentally testable.

Lajos Diósi \cite{4,5,6,7} and Roger Penrose \cite{8,9,10} proposed one such objective theory. They argue that the superposition of the wavefunction is at odds with general relativity because it results in two incompatible spacetimes. In Schrödinger's thought experiment, one might imagine the live cat remaining standing while the dead cat falls to the floor, giving spacetime two different curvatures. The paradox must end when the incompatibility exceeds the time-energy version of the Heisenberg uncertainty principle. Penrose and Diósi derive the equation for the collapse in two ways, differing only by a factor of two in the self-energy gamma ($\gamma$) constant; hence, their interpretation is called the Diósi-Penrose model \cite{11}.

\begin{figure}[ht]
    \centering
    \includegraphics[width=0.48\textwidth]{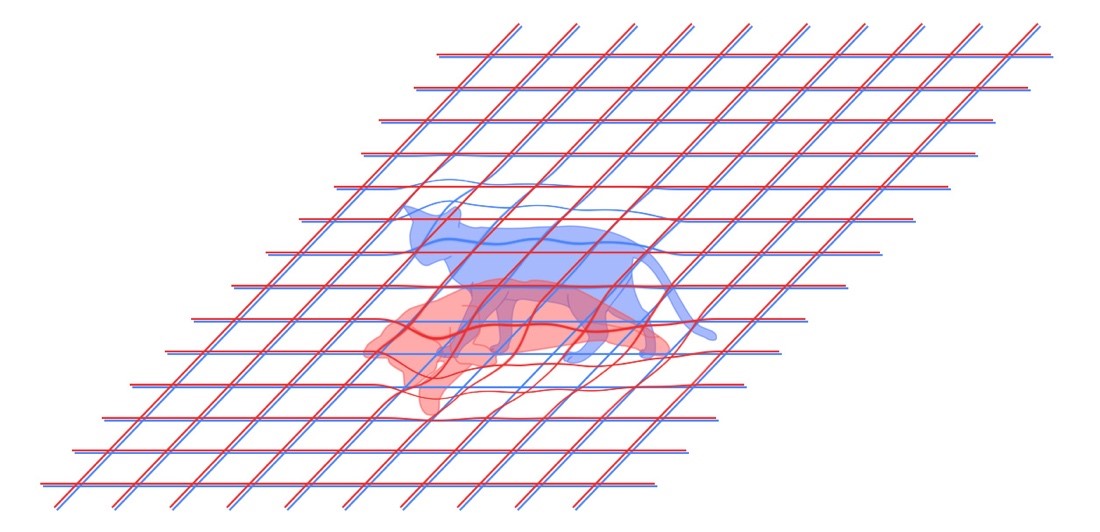}
    \caption{Superposition of massive quantum objects, for example, a live cat standing (blue) and a dead cat laying down (red), results in two different configurations of spacetime. In the Diósi-Penrose interpretation once the difference in spacetime exceeds the time-energy uncertainty limit, the wavefunction must collapse to avoid a paradox.}
    \label{fig:1}
\end{figure}

In 2020, an experiment \cite{12} was performed in a salt mine with a negative result, and it was suggested this refuted the Diósi-Penrose interpretation. Although the Diósi and Penrose results are similar, they differ in some important details. The Diósi mechanism requires the delocalized wavefunction to reunite upon collapse. Researchers pointed out that if this particle were charged, it should reveal itself through the emission of an electromagnetic signature. The salt mine experiment looked for the characteristic radiation and found none. However, Penrose does not propose this mechanism but rather a purer mathematical solution. To prevent a paradoxical spacetime, an effect propagates back in time to cause the photon to have taken one or another path, and the second path is erased from the timeline. He does not attempt to explain the mechanism of the backward time signaling, but since there is no reuniting, no electromagnetic signature would be emitted. We propose an experiment to directly observe the collapse of the wavefunction in the Penrose interpretation.

First, a thought experiment. In Young's double-slit experiment, a light is shone through two slits and interferes with itself to produce the classic double-slit pattern. What would happen if we covered those slits with shutters that open in superposition? If collapse is instantaneous, then only one shutter would ever be open, and we would see a single line with faint knife-edge diffraction. But, if both shutters were to open, we would see the classic Young’s double slit interference pattern for a brief period until collapse occurred. Up until now, the electronics to build such an apparatus would involve too much mass, but the advent of thin-film solid state devices such as single photon avalanche diodes (SPADs) makes it possible to build experiments with very small effective mass. We are not concerned with the overall mass of the device, just the mass that would be in two places at once. Minimizing the total effective mass reveals a testable difference between interpretations that rely on ‘which-way’ information and the Diósi-Penrose model, which depends on spacetime mass distribution.

In the thought experiment, a source of single photons hits a beam splitter which in turn triggers two SPADs. The current from each SPAD controls a piezo motor that opens one of the shutters. Do the shutters open in superposition, or does the SPAD collapse the wavefunction? The SPAD clearly has sufficient which-way information to act as an observer. It ‘knows’ the path the photon took. But is the SPAD the measuring device, or must we consider the combined system of SPAD, piezo, and shutter? Partitioning the system is subjective and dependent on human labels. Could one person label the experiment one way and another a different way to get different results? This reminds us of John Bell’s quip that we might have to wait for observation by a physicist with a PhD to collapse the wavefunction. 

Considering the entire system as the measuring apparatus sidesteps the subjective labeling problem. Collapse is driven by the difference in mass distribution. It is still affected by an observer but in an objective manner. If a physicist observes the experiment and the experiment is in superposition, the mass of the physicist is added to the superposition. At first, the addition is minimal. A few molecules in their retina are superposed. After a few moments, many neurons are involved, and eventually, they might put up their hand and say, “I saw the North detector fire.” Certainly, the motion of any part of their body puts sufficient mass into superposition to instantly cause collapse. It is interesting to ask where in the chain of thoughts, from the retina to action, the collapse is triggered. The trick is to make the entire apparatus sufficiently small that we can watch the evolution of the interference pattern and observe collapse in real time.

\begin{figure}[ht]
    \centering
    \includegraphics[width=0.48\textwidth]{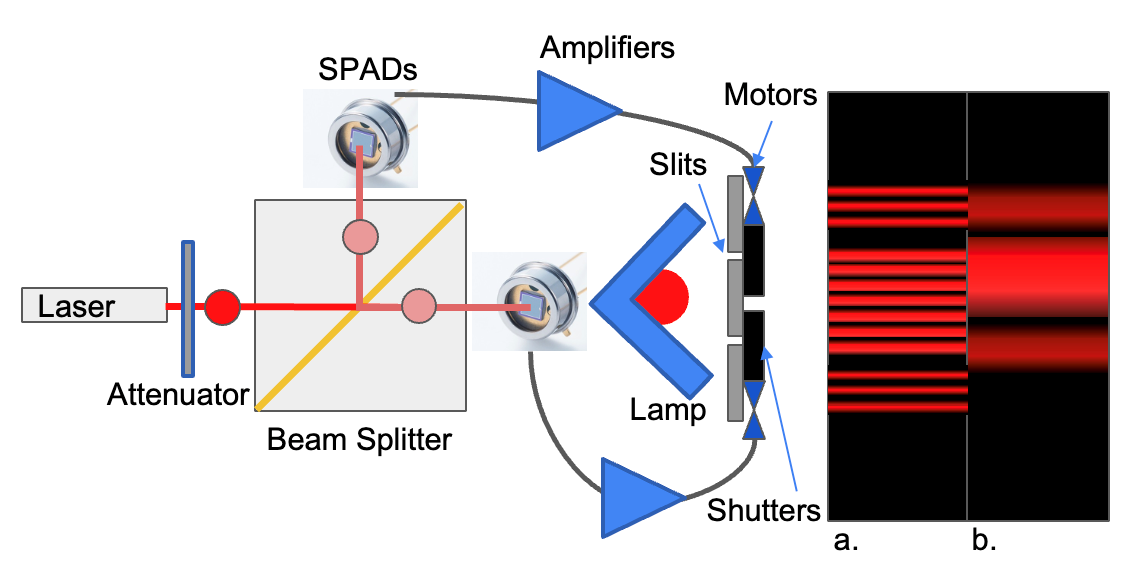}
    \caption{A thought experiment to detect the collapse of the wavefunction using Young’s double slit apparatus. A laser is attenuated to provide a stream of single photons which trigger two SPADs. Each SPAD controls a motor through an amplifier which opens a shutter. If the mass of the electrons involved in detection and amplification and the mass of the motors and shutters is sufficiently low, you should see a momentary double slit interference pattern – a, followed by a collapse to the single slit pattern – b.}
    \label{fig:2}
\end{figure}

\section{Collapse Time.}

The Heisenberg time-energy uncertainty equation \eqref{eq:1} determines the time for collapse. The larger the energy, the faster the collapse.
\begin{equation}\label{eq:1}
 t=\gamma \frac{\hbar}{E_g} 
\end{equation}

Where $E_g$ is the gravitational self-energy, $\hbar$ is the reduced Plank constant, and $\gamma$ is a constant originally estimated by Penrose to be $1/(8\pi)$.

This is the same energy released when dust clouds coalesce and ‘fall’ into the gravitational well to form a planet. If we wish to put our planet into superposition, we must work against this enormous energy. We should immediately say that Penrose is \textit{not} arguing we need to find this energy. That would conflict with conservation laws. Instead, when it becomes certain we \textit{need} to find this energy, the wavefunction collapses instead.

This time dependence solves Schrödinger’s paradox: the wavefunction of a proton separated by its radius would collapse in $10^{7}$ years, a dust particle would collapse in $10^{-8}$ seconds, and a cat collapses in $10^{-28}$ seconds \cite{13}. We don’t see superpositions of alive and dead cats because they are far too short-lived for our visual perception to register.

In the Howl, Penrose, Fuentes (HPF) paper \cite{14}, the energy is calculated for two regions: large separation and small separation determined by \textit{Lambda} ($\lambda$) - the ratio of the separation distance $s$ to the size of the separated mass $R$;
\begin{equation}\label{eq:2}
\lambda = \frac{\Delta s}{2R} 
\end{equation}

For a small mass separated by a large distance, the region where ($\lambda \geq 1$), the self-energy is given by;
\begin{equation}\label{eq:3}
 Eg = \frac{6GM^2}{5R} \left(1 - \frac{5}{12\lambda}\right)
 \end{equation}

Where $G$ is the gravitational constant, $M$ is the mass in superposition, and $R$ is the radius of that mass – assuming it is a sphere. For a large mass separated by a small distance, the region where ($0\leq \lambda \leq 1$), the self-energy is given by;
\begin{equation}\label{eq:4}
E_g = \frac{6GM^2}{5R} \left(\frac{5}{3} \lambda^2 - \frac{5}{3} \lambda^3 + \frac{1}{6} \lambda^5 \right)
\end{equation}

We can see for small $\lambda$, collapse time varies with the square of the separation. Thus, we can use mesoscopic masses provided we move them a very small distance.

\vspace{10pt}
\section{The Experiment.}

The mechanical shutters in our thought experiment are too massive in practice and must also be moved a large distance to open and close the slits. To overcome this, we propose a Mach-Zehnder interferometer with small corner mirrors mounted on piezoelectric actuators that can move a few microns. There might be other solutions, such as lightweight LCD shutters or a directly modulated light valve, but the Mach-Zehnder is readily constructed with standard laboratory equipment. 

A first laser is used as a source of single photons to trigger the SPADs in superposition. The beam from a second laser is split by a beam splitter and hits two mirrors. Each mirror is mounted on a piezo actuator controlled by a SPAD. The beams are recombined and sent to a detector. If either mirror moves, the interference pattern is altered, and the intensity of the light at the detector changes. If both mirrors move equally, the two paths change by the same length and no change in the pattern should be seen. If collapse takes time, we should see a delay before the interference pattern changes. The apparatus must detect the change in interference pattern representing one Angstrom unit in better than 100 nanoseconds. The lightest readily available commercial mirror weighs approximately 0.2 grams.

Garrelt Quandt-Wiese had earlier calculated gravitational collapse times for several practical electronic components including SPADs \cite{15}. The method is slightly simpler than the HPF paper but gives similar answers. SPADs, for example, would take 1000 seconds for the wavefunction to collapse, while copper wiring takes $10^{10}$ seconds, resistors take 50 seconds and piezo components around 100 milliseconds. The large collapse times are counterintuitive, but consider the only masses involved are electrons and small movements of the silicon and copper latices through self-heating. This leaves the mirrors as the main contributors to the collapse time at around 1.6 microseconds. This is broadly in agreement with the analysis of an oblate sphere using the HPF paper, which puts the collapse time for our mirrors at around 1.4 microseconds.

\begin{figure}[ht]
    \centering
    \includegraphics[width=0.48\textwidth]{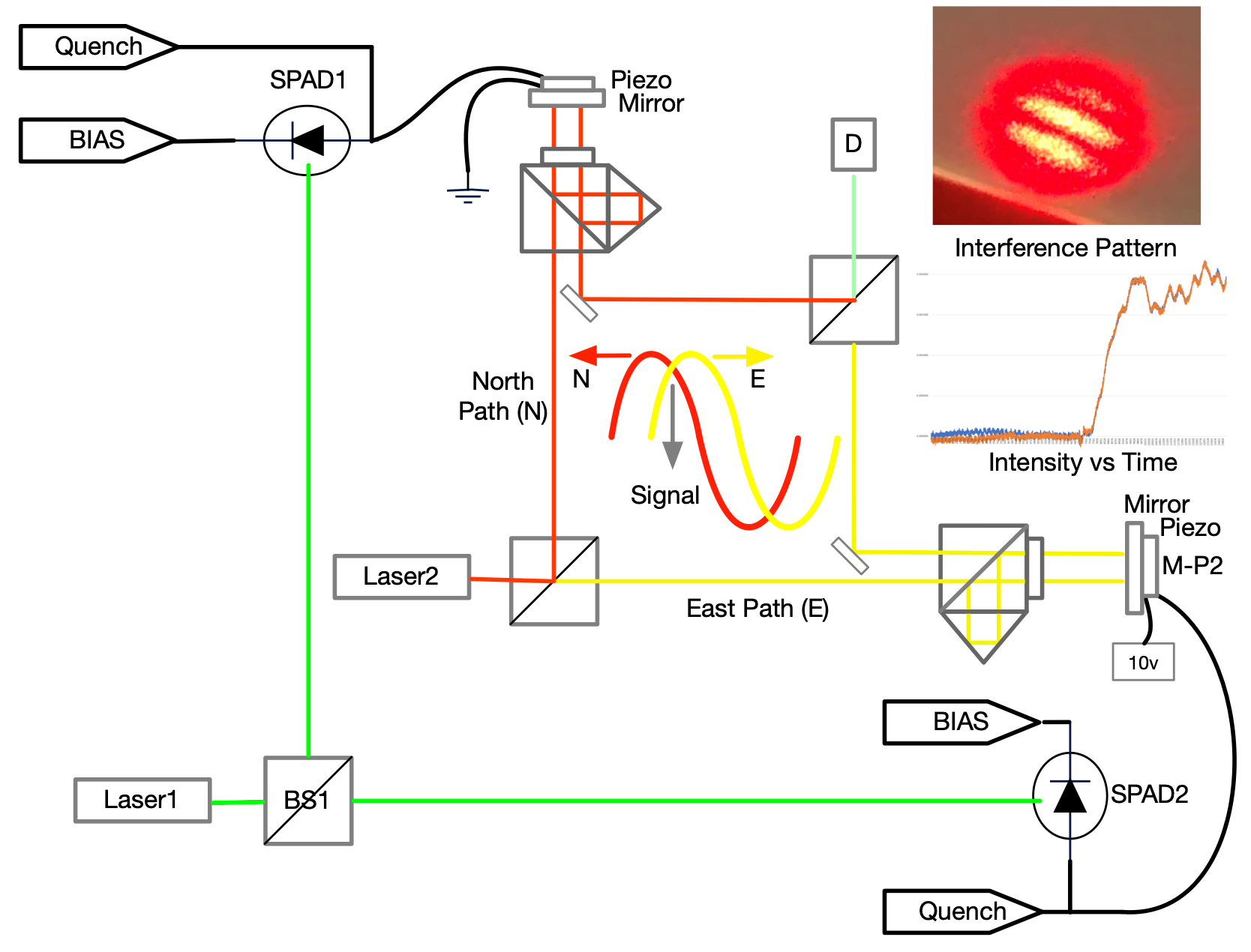}
    \caption{A Mach-Zehnder interferometer has its path lengths varied by two piezo-controlled mirrors. Single-photon avalanche diodes (SPADs) connected to the output of a first beam splitter control these mirrors. If the mirrors move in superposition, the interference pattern will not change until collapse occurs.}
    \label{fig:3}
\end{figure}

There are several issues to handle: minimizing the effective mass, coping with noise, and avoiding leakage of ‘which-way’ information. The Diósi-Penrose interpretation does not believe which-way information per se is affecting collapse. It is still a convenient shorthand, meaning amplification of a difference in mass distribution to the threshold for collapse. A human observer is one such amplifier, but other more subtle ones occur throughout the system and must be avoided. Perhaps we should use the term which-way difference rather than which-way information. Single photon detectors contain quench circuits to re-trigger the system. We can’t use them as the re-triggering circuit gains which-way information. We must also be careful that modern power supplies contain microprocessors to stabilize their output. These could gain which-way information, so any power supply must be configured to feed the two SPADs equally so there is no potential for the power supplies to detect which SPAD has triggered.

There is a worry that the mounting for the piezo SPAD assembly must also be considered in the effective mass. However, here conservation of momentum comes to our rescue. The mirror mounts weigh 20 grams while the mirrors only 0.2 grams. If the mirror moves one micron, the mount moves 1/100\textsuperscript{th} of a micron. Because we are in inverse square law region of the self-energy calculation, the mount will contribute 1/10,000\textsuperscript{th} due to its motion and therefore 1/100\textsuperscript{th} of the self-energy. There is a brief time as the equal and opposite reaction spreads out into the material of the mount at the speed of sound, but this will only add 10\% to the collapse time at most.

All the normal techniques to reduce noise should be employed: housing the apparatus in a Faraday cage, isolating the optical components on an air table, and cooling the SPAD at least 30 degrees below ambient conditions. This reduces dark count by a factor of 1000 to around 10,000 dark counts per second. Dark count can be used to our advantage, offering a control as it is never a superposed event. Turning the first laser on and off gives us the ability to run with superposition on or off. We can then plot the intensity of the interference pattern against time for the two cases and see whether they differ. If they differ we have evidence that collapse of the wavefunction is taking time. We can further use the intensity to measure displacement and estimate $\gamma$ in the Diósi-Penrose equation.

There is a potential problem with the experimental setup. When the interference pattern changes, it reveals which-way information about the photon. The measurement is different for a North-going and East-going photon: Increasing amplitude in one case and decreasing in the other. This difference is initially minor but would rapidly entrain the mass of the oscilloscope and, ultimately, the experimenter, causing rapid collapse. How rapid is unclear. It could operate as soon as \textit{any} which-way difference is available curtailing the experiment almost instantaneously. Or worse still it could refer back in time and negate the experiment entirely. We need to erase this which-way difference \cite{16}. We can achieve this by reversing the motion of one of the mirrors. The experiment can be run with and without this quantum eraser. A further eraser is needed to trigger the oscilloscope from the output of the two single-photon detectors. A simple summing junction will suffice. This allows us to observe the onset of the experiment without the observation causing collapse. We know a SPAD fired but don’t know which one. Thus, we can eliminate all leakage of which-way information or which-way differences in the configuration of our apparatus and concentration on variations in the the two mirrors.

\vspace{10pt}
\section{The Decoherence Problem.}

The mirrors consist of $10^{21}$ atoms each, six orders of magnitude larger than the quantum drums, which hold the current record for macroscopic superposition \cite{17}. We don’t claim to be breaking the record for putting devices into superposition in the normal sense because our devices are at room temperature in ambient conditions, and their coherence breaks down in an instant. But, we can still sense collapse time even though each mirror has decohered. Phase information is delicate, and the room-temperature decoherence time for our mirrors is on the order of $10^{-38}$ seconds \cite{18}. Even with cryogenic refrigeration and ultra-high vacuum, our mirrors would decohere in the order of $10^{-20}$ seconds. Many scientists believe decoherence and collapse are synonymous, but this is a misconception \cite{19,20}. Decoherence usually overwhelms attempts to observe gravitational collapse, but we can observe collapse independently of decoherence in our symmetrical system. The mathematics explains this. The density matrix for the two mirror systems activated by a single photon is:
\begin{equation}\label{eq:5}
\begin{pmatrix}
\frac{1}{2} & \frac{e^{-i\theta}}{2}\\
\frac{e^{i\theta}}{2} & \frac{1}{2}
\end{pmatrix}
\end{equation}

This system is in a 50:50 superposition, and phase information is in the off-diagonal terms. As air molecules buffet the mirrors, the off-diagonal terms tend to zero. The density matrix transitions to:
\begin{equation}\label{eq:6}
\begin{pmatrix}
\frac{1}{2} & \to 0 \\ 
\to 0 & \frac{1}{2} 
\end{pmatrix}
\end{equation}

Taking the mathematics literally, we now have a matrix that is neither quantum nor classical. No wavefunction corresponds to this density matrix, nor is this a classical state. For collapse to occur, the density matrix of our experiment must break symmetry and transition to either:
\begin{equation}\label{eq:7}
\begin{pmatrix}
1 & 0 \\ 
0 & 0
\end{pmatrix} \ or \
\begin{pmatrix}
0 & 0 \\ 
0 & 1 \\
\end{pmatrix}
\end{equation}

The separation of two quantum features in an experiment – decoherence and collapse in our case – is not without precedent. In the Cheshire Cat experiment \cite{21,22}, polarization and localization are independently measured, although in a recent paper, this has been refuted \cite{23}. We believe something analogous is happening here and that we can separate decoherence – a noise-like process affecting phase information that needs cryogenic refrigeration to suppress – from gravitational collapse – a symmetry-breaking process affecting localization which is unaffected by temperature. Perhaps we can rehabilitate the Cheshire Cat.

\vspace{10pt}
\section{Conclusion.}

A window exists where an appropriate choice of mass and displacement should allow observation of wavefunction collapse in ambient laboratory conditions. Using a symmetrical experiment separates decoherence of phase information from collapse of the localization allowing the properties to be independently measured. The effect has application in quantum computing where it may allow an increase in computational power over conventional quantum computers \cite{24,25}, communications systems, and sensors. The experiment is compatible with observer models in the high mass limit, but shows that observation must be understood as an amplification process that takes time. The proposal is also compatible with a many-worlds view. The Heisenberg time-energy limit is probabilistic. That means there is a finite possibility that some wavefunctions will never collapse. The only way to prune these multi-verses to zero is if collapse is quantized. We hope our experiment will be able to probe such issues.

\section{Data Availability.}
Patent WO2024020006 - QUANTUM GRAVITY DEVICE gives detailed instructions for constructing our experimental apparatus.

\section{Acknowledgments.}
Thanks to Garrelt Quandt-Wiese, for the calculations supporting this experiment, to XCOM-Labs for providing lab space, to the UCSD physics department for their encouragement and offer of facilities that prompted us to begin this experiment, and to Jen Root, who provided a gift to purchase equipment. Due to Covid restrictions, initial experiments were performed in the Tagg family garage – for which we must thank Kelly, Bianca, and Imogen for their patience.

%\nocite{*}

\bibliography{bibliography}% Produces the bibliography via BibTeX.

\providecommand{\noopsort}[1]{}\providecommand{\singleletter}[1]{#1}%
\begin{thebibliography}{25}%
\makeatletter
\providecommand \@ifxundefined [1]{%
 \@ifx{#1\undefined}
}%
\providecommand \@ifnum [1]{%
 \ifnum #1\expandafter \@firstoftwo
 \else \expandafter \@secondoftwo
 \fi
}%
\providecommand \@ifx [1]{%
 \ifx #1\expandafter \@firstoftwo
 \else \expandafter \@secondoftwo
 \fi
}%
\providecommand \natexlab [1]{#1}%
\providecommand \enquote  [1]{``#1''}%
\providecommand \bibnamefont  [1]{#1}%
\providecommand \bibfnamefont [1]{#1}%
\providecommand \citenamefont [1]{#1}%
\providecommand \href@noop [0]{\@secondoftwo}%
\providecommand \href [0]{\begingroup \@sanitize@url \@href}%
\providecommand \@href[1]{\@@startlink{#1}\@@href}%
\providecommand \@@href[1]{\endgroup#1\@@endlink}%
\providecommand \@sanitize@url [0]{\catcode `\\12\catcode `\$12\catcode `\&12\catcode `\#12\catcode `\^12\catcode `\_12\catcode `\%12\relax}%
\providecommand \@@startlink[1]{}%
\providecommand \@@endlink[0]{}%
\providecommand \url  [0]{\begingroup\@sanitize@url \@url }%
\providecommand \@url [1]{\endgroup\@href {#1}{\urlprefix }}%
\providecommand \urlprefix  [0]{URL }%
\providecommand \Eprint [0]{\href }%
\providecommand \doibase [0]{https://doi.org/}%
\providecommand \selectlanguage [0]{\@gobble}%
\providecommand \bibinfo  [0]{\@secondoftwo}%
\providecommand \bibfield  [0]{\@secondoftwo}%
\providecommand \translation [1]{[#1]}%
\providecommand \BibitemOpen [0]{}%
\providecommand \bibitemStop [0]{}%
\providecommand \bibitemNoStop [0]{.\EOS\space}%
\providecommand \EOS [0]{\spacefactor3000\relax}%
\providecommand \BibitemShut  [1]{\csname bibitem#1\endcsname}%
\let\auto@bib@innerbib\@empty
%</preamble>
\bibitem [{\citenamefont {Schr{\"o}dinger}(1935)}]{1}%
  \BibitemOpen
  \bibfield  {author} {\bibinfo {author} {\bibfnamefont {E.}~\bibnamefont {Schr{\"o}dinger}},\ }\bibfield  {title} {\bibinfo {title} {Die gegenw{\"a}rtige situation in der quantenmechanik},\ }\href {https://doi.org/10.1007/BF01491891} {\bibfield  {journal} {\bibinfo  {journal} {Naturwissenschaften}\ }\textbf {\bibinfo {volume} {23}},\ \bibinfo {pages} {807} (\bibinfo {year} {1935})}\BibitemShut {NoStop}%
\bibitem [{\citenamefont {Wigner}(1961)}]{2}%
  \BibitemOpen
  \bibfield  {author} {\bibinfo {author} {\bibfnamefont {E.~P.}\ \bibnamefont {Wigner}},\ }\bibfield  {title} {\bibinfo {title} {Remarks on the mind-body question},\ }in\ \href@noop {} {\emph {\bibinfo {booktitle} {The Scientist Speculates}}},\ \bibinfo {editor} {edited by\ \bibinfo {editor} {\bibfnamefont {I.~J.}\ \bibnamefont {Good}}}\ (\bibinfo  {publisher} {Springer},\ \bibinfo {year} {1961})\ pp.\ \bibinfo {pages} {247--260}\BibitemShut {NoStop}%
\bibitem [{\citenamefont {Everett~III}(1957)}]{3}%
  \BibitemOpen
  \bibfield  {author} {\bibinfo {author} {\bibfnamefont {H.}~\bibnamefont {Everett~III}},\ }\bibfield  {title} {\bibinfo {title} {"relative state" formulation of quantum mechanics},\ }\href {https://doi.org/10.1103/RevModPhys.29.454} {\bibfield  {journal} {\bibinfo  {journal} {Reviews of Modern Physics}\ }\textbf {\bibinfo {volume} {29}},\ \bibinfo {pages} {454} (\bibinfo {year} {1957})}\BibitemShut {NoStop}%
\bibitem [{\citenamefont {Di{\'o}si}(2005)}]{4}%
  \BibitemOpen
  \bibfield  {author} {\bibinfo {author} {\bibfnamefont {L.}~\bibnamefont {Di{\'o}si}},\ }\bibfield  {title} {\bibinfo {title} {Intrinsic time-uncertainties and decoherence: comparison of 4 models},\ }\href {https://doi.org/10.1590/S0103-97332005000200009} {\bibfield  {journal} {\bibinfo  {journal} {Brazilian Journal of Physics}\ }\textbf {\bibinfo {volume} {35}},\ \bibinfo {pages} {260} (\bibinfo {year} {2005})}\BibitemShut {NoStop}%
\bibitem [{\citenamefont {Diosi}(1984)}]{5}%
  \BibitemOpen
  \bibfield  {author} {\bibinfo {author} {\bibfnamefont {L.}~\bibnamefont {Diosi}},\ }\bibfield  {title} {\bibinfo {title} {Gravitation and quantum-mechanical localization of macro-objects},\ }\bibfield  {journal} {\bibinfo  {journal} {Physics Letters A}\ }\textbf {\bibinfo {volume} {105}},\ \href {https://doi.org/10.1016/0375-9601(84)90397-9} {10.1016/0375-9601(84)90397-9} (\bibinfo {year} {1984})\BibitemShut {NoStop}%
\bibitem [{\citenamefont {Diosi}(1987)}]{6}%
  \BibitemOpen
  \bibfield  {author} {\bibinfo {author} {\bibfnamefont {L.}~\bibnamefont {Diosi}},\ }\bibfield  {title} {\bibinfo {title} {A universal master equation for the gravitational violation of quantum mechanics},\ }\href {https://doi.org/10.1016/0375-9601(87)90681-5} {\bibfield  {journal} {\bibinfo  {journal} {Physics letters A}\ }\textbf {\bibinfo {volume} {120}},\ \bibinfo {pages} {377} (\bibinfo {year} {1987})}\BibitemShut {NoStop}%
\bibitem [{\citenamefont {Di{\'o}si}(1989)}]{7}%
  \BibitemOpen
  \bibfield  {author} {\bibinfo {author} {\bibfnamefont {L.}~\bibnamefont {Di{\'o}si}},\ }\bibfield  {title} {\bibinfo {title} {Models for universal reduction of macroscopic quantum fluctuations},\ }\href {https://doi.org/10.1103/PhysRevA.40.1165} {\bibfield  {journal} {\bibinfo  {journal} {Physical Review A}\ }\textbf {\bibinfo {volume} {40}},\ \bibinfo {pages} {1165} (\bibinfo {year} {1989})}\BibitemShut {NoStop}%
\bibitem [{\citenamefont {Penrose}(2014)}]{8}%
  \BibitemOpen
  \bibfield  {author} {\bibinfo {author} {\bibfnamefont {R.}~\bibnamefont {Penrose}},\ }\bibfield  {title} {\bibinfo {title} {On the gravitization of quantum mechanics 1: Quantum state reduction},\ }\href {https://doi.org/10.1007/s10701-013-9770-0} {\bibfield  {journal} {\bibinfo  {journal} {Foundations of Physics}\ }\textbf {\bibinfo {volume} {44}},\ \bibinfo {pages} {557} (\bibinfo {year} {2014})}\BibitemShut {NoStop}%
\bibitem [{\citenamefont {Penrose}\ and\ \citenamefont {Isham}(1986)}]{9}%
  \BibitemOpen
  \bibfield  {author} {\bibinfo {author} {\bibfnamefont {R.}~\bibnamefont {Penrose}}\ and\ \bibinfo {author} {\bibfnamefont {C.~J.}\ \bibnamefont {Isham}},\ }\href@noop {} {\bibinfo {title} {Quantum concepts in space and time}} (\bibinfo {year} {1986})\BibitemShut {NoStop}%
\bibitem [{\citenamefont {Penrose}(1996)}]{10}%
  \BibitemOpen
  \bibfield  {author} {\bibinfo {author} {\bibfnamefont {R.}~\bibnamefont {Penrose}},\ }\bibfield  {title} {\bibinfo {title} {On gravity's role in quantum state reduction},\ }\href {https://doi.org/10.1007/BF02105068} {\bibfield  {journal} {\bibinfo  {journal} {General Relativity and Gravitation}\ }\textbf {\bibinfo {volume} {28}},\ \bibinfo {pages} {581} (\bibinfo {year} {1996})}\BibitemShut {NoStop}%
\bibitem [{\citenamefont {Bassi}(2016)}]{11}%
  \BibitemOpen
  \bibfield  {author} {\bibinfo {author} {\bibfnamefont {A.}~\bibnamefont {Bassi}},\ }\bibfield  {title} {\bibinfo {title} {Models of spontaneous wave function collapse: what they are, and how they can be tested},\ }in\ \href {https://doi.org/10.1088/1742-6596/701/1/012012} {\emph {\bibinfo {booktitle} {Journal of Physics: Conference Series}}},\ Vol.\ \bibinfo {volume} {701}\ (\bibinfo {organization} {IOP Publishing},\ \bibinfo {year} {2016})\ p.\ \bibinfo {pages} {012012}\BibitemShut {NoStop}%
\bibitem [{\citenamefont {Donadi}\ \emph {et~al.}(2021)\citenamefont {Donadi}, \citenamefont {Piscicchia}, \citenamefont {Curceanu}, \citenamefont {Di{\'o}si}, \citenamefont {Laubenstein},\ and\ \citenamefont {Bassi}}]{12}%
  \BibitemOpen
  \bibfield  {author} {\bibinfo {author} {\bibfnamefont {S.}~\bibnamefont {Donadi}}, \bibinfo {author} {\bibfnamefont {K.}~\bibnamefont {Piscicchia}}, \bibinfo {author} {\bibfnamefont {C.}~\bibnamefont {Curceanu}}, \bibinfo {author} {\bibfnamefont {L.}~\bibnamefont {Di{\'o}si}}, \bibinfo {author} {\bibfnamefont {M.}~\bibnamefont {Laubenstein}},\ and\ \bibinfo {author} {\bibfnamefont {A.}~\bibnamefont {Bassi}},\ }\bibfield  {title} {\bibinfo {title} {Underground test of gravity-related wave function collapse},\ }\href {https://doi.org/10.1038/s41567-020-1008-4} {\bibfield  {journal} {\bibinfo  {journal} {Nature Physics}\ }\textbf {\bibinfo {volume} {17}},\ \bibinfo {pages} {74} (\bibinfo {year} {2021})}\BibitemShut {NoStop}%
\bibitem [{\citenamefont {Schlosshauer}(2019{\natexlab{a}})}]{13}%
  \BibitemOpen
  \bibfield  {author} {\bibinfo {author} {\bibfnamefont {M.}~\bibnamefont {Schlosshauer}},\ }\bibfield  {title} {\bibinfo {title} {The quantum-to-classical transition and decoherence},\ }\href {http://arxiv.org/abs/1404.2635} {\bibfield  {journal} {\bibinfo  {journal} {arXiv preprint arXiv:1404.2635}\ } (\bibinfo {year} {2019}{\natexlab{a}})}\BibitemShut {NoStop}%
\bibitem [{\citenamefont {Howl}\ \emph {et~al.}(2019)\citenamefont {Howl}, \citenamefont {Penrose},\ and\ \citenamefont {Fuentes}}]{14}%
  \BibitemOpen
  \bibfield  {author} {\bibinfo {author} {\bibfnamefont {R.}~\bibnamefont {Howl}}, \bibinfo {author} {\bibfnamefont {R.}~\bibnamefont {Penrose}},\ and\ \bibinfo {author} {\bibfnamefont {I.}~\bibnamefont {Fuentes}},\ }\bibfield  {title} {\bibinfo {title} {Exploring the unification of quantum theory and general relativity with a bose--einstein condensate},\ }\href {https://doi.org/10.1088/1367-2630/ab104a} {\bibfield  {journal} {\bibinfo  {journal} {New Journal of Physics}\ }\textbf {\bibinfo {volume} {21}},\ \bibinfo {pages} {043047} (\bibinfo {year} {2019})}\BibitemShut {NoStop}%
\bibitem [{\citenamefont {Quandt-Wiese}(2017)}]{15}%
  \BibitemOpen
  \bibfield  {author} {\bibinfo {author} {\bibfnamefont {G.}~\bibnamefont {Quandt-Wiese}},\ }\bibfield  {title} {\bibinfo {title} {Di{\'o}si-penrose criterion for solids in quantum superpositions and a single-photon detector},\ }\href {http://arxiv.org/abs/1701.00353} {\bibfield  {journal} {\bibinfo  {journal} {arXiv preprint arXiv:1701.00353}\ } (\bibinfo {year} {2017})}\BibitemShut {NoStop}%
\bibitem [{\citenamefont {Kim}\ \emph {et~al.}(2000)\citenamefont {Kim}, \citenamefont {Yu}, \citenamefont {Kulik}, \citenamefont {Shih},\ and\ \citenamefont {Scully}}]{16}%
  \BibitemOpen
  \bibfield  {author} {\bibinfo {author} {\bibfnamefont {Y.-H.}\ \bibnamefont {Kim}}, \bibinfo {author} {\bibfnamefont {R.}~\bibnamefont {Yu}}, \bibinfo {author} {\bibfnamefont {S.~P.}\ \bibnamefont {Kulik}}, \bibinfo {author} {\bibfnamefont {Y.}~\bibnamefont {Shih}},\ and\ \bibinfo {author} {\bibfnamefont {M.~O.}\ \bibnamefont {Scully}},\ }\bibfield  {title} {\bibinfo {title} {Delayed “choice” quantum eraser},\ }\href {https://doi.org/10.1103/PhysRevLett.84.1} {\bibfield  {journal} {\bibinfo  {journal} {Physical Review Letters}\ }\textbf {\bibinfo {volume} {84}},\ \bibinfo {pages} {1} (\bibinfo {year} {2000})}\BibitemShut {NoStop}%
\bibitem [{\citenamefont {Castelvecchi}(2021)}]{17}%
  \BibitemOpen
  \bibfield  {author} {\bibinfo {author} {\bibfnamefont {D.}~\bibnamefont {Castelvecchi}},\ }\bibfield  {title} {\bibinfo {title} {Minuscule drums push the limits of quantum weirdness},\ }\bibfield  {journal} {\bibinfo  {journal} {Nature}\ }\href {https://doi.org/10.1038/d41586-021-01223-4} {10.1038/d41586-021-01223-4} (\bibinfo {year} {2021})\BibitemShut {NoStop}%
\bibitem [{\citenamefont {Schlosshauer}(2005)}]{18}%
  \BibitemOpen
  \bibfield  {author} {\bibinfo {author} {\bibfnamefont {M.}~\bibnamefont {Schlosshauer}},\ }\bibfield  {title} {\bibinfo {title} {Decoherence, the measurement problem, and interpretations of quantum mechanics},\ }\href {https://doi.org/10.1103/RevModPhys.76.1267} {\bibfield  {journal} {\bibinfo  {journal} {Reviews of Modern physics}\ }\textbf {\bibinfo {volume} {76}},\ \bibinfo {pages} {1267} (\bibinfo {year} {2005})}\BibitemShut {NoStop}%
\bibitem [{\citenamefont {Sivasundaram}\ and\ \citenamefont {Nielsen}(2016)}]{19}%
  \BibitemOpen
  \bibfield  {author} {\bibinfo {author} {\bibfnamefont {S.}~\bibnamefont {Sivasundaram}}\ and\ \bibinfo {author} {\bibfnamefont {K.~H.}\ \bibnamefont {Nielsen}},\ }\bibfield  {title} {\bibinfo {title} {Surveying the attitudes of physicists concerning foundational issues of quantum mechanics},\ }\bibfield  {journal} {\bibinfo  {journal} {arXiv preprint arXiv:1612.00676}\ }\href {https://doi.org/10.48550/arXiv.1612.00676} {10.48550/arXiv.1612.00676} (\bibinfo {year} {2016})\BibitemShut {NoStop}%
\bibitem [{\citenamefont {Schlosshauer}(2019{\natexlab{b}})}]{20}%
  \BibitemOpen
  \bibfield  {author} {\bibinfo {author} {\bibfnamefont {M.}~\bibnamefont {Schlosshauer}},\ }\bibfield  {title} {\bibinfo {title} {Quantum decoherence},\ }\href {https://doi.org/10.1016/j.physrep.2019.10.001} {\bibfield  {journal} {\bibinfo  {journal} {Physics Reports}\ }\textbf {\bibinfo {volume} {831}},\ \bibinfo {pages} {1} (\bibinfo {year} {2019}{\natexlab{b}})}\BibitemShut {NoStop}%
\bibitem [{\citenamefont {Duprey}\ \emph {et~al.}(2018)\citenamefont {Duprey}, \citenamefont {Kanjilal}, \citenamefont {Sinha}, \citenamefont {Home},\ and\ \citenamefont {Matzkin}}]{21}%
  \BibitemOpen
  \bibfield  {author} {\bibinfo {author} {\bibfnamefont {Q.}~\bibnamefont {Duprey}}, \bibinfo {author} {\bibfnamefont {S.}~\bibnamefont {Kanjilal}}, \bibinfo {author} {\bibfnamefont {U.}~\bibnamefont {Sinha}}, \bibinfo {author} {\bibfnamefont {D.}~\bibnamefont {Home}},\ and\ \bibinfo {author} {\bibfnamefont {A.}~\bibnamefont {Matzkin}},\ }\bibfield  {title} {\bibinfo {title} {The quantum cheshire cat effect: theoretical basis and observational implications},\ }\href {https://doi.org/10.1016/j.aop.2018.01.011} {\bibfield  {journal} {\bibinfo  {journal} {Annals of Physics}\ }\textbf {\bibinfo {volume} {391}},\ \bibinfo {pages} {1} (\bibinfo {year} {2018})}\BibitemShut {NoStop}%
\bibitem [{\citenamefont {Aharonov}\ \emph {et~al.}(2013)\citenamefont {Aharonov}, \citenamefont {Popescu}, \citenamefont {Rohrlich},\ and\ \citenamefont {Skrzypczyk}}]{22}%
  \BibitemOpen
  \bibfield  {author} {\bibinfo {author} {\bibfnamefont {Y.}~\bibnamefont {Aharonov}}, \bibinfo {author} {\bibfnamefont {S.}~\bibnamefont {Popescu}}, \bibinfo {author} {\bibfnamefont {D.}~\bibnamefont {Rohrlich}},\ and\ \bibinfo {author} {\bibfnamefont {P.}~\bibnamefont {Skrzypczyk}},\ }\bibfield  {title} {\bibinfo {title} {Quantum cheshire cats},\ }\href {https://doi.org/10.1088/1367-2630/15/11/113015} {\bibfield  {journal} {\bibinfo  {journal} {New Journal of Physics}\ }\textbf {\bibinfo {volume} {15}},\ \bibinfo {pages} {113015} (\bibinfo {year} {2013})}\BibitemShut {NoStop}%
\bibitem [{\citenamefont {Hance}\ \emph {et~al.}(2023)\citenamefont {Hance}, \citenamefont {Ji},\ and\ \citenamefont {Hofmann}}]{23}%
  \BibitemOpen
  \bibfield  {author} {\bibinfo {author} {\bibfnamefont {J.~R.}\ \bibnamefont {Hance}}, \bibinfo {author} {\bibfnamefont {M.}~\bibnamefont {Ji}},\ and\ \bibinfo {author} {\bibfnamefont {H.~F.}\ \bibnamefont {Hofmann}},\ }\bibfield  {title} {\bibinfo {title} {Contextuality, coherences, and quantum cheshire cats},\ }\href {https://doi.org/10.1088/1367-2630/ad0bd4} {\bibfield  {journal} {\bibinfo  {journal} {New Journal of Physics}\ }\textbf {\bibinfo {volume} {25}},\ \bibinfo {pages} {113028} (\bibinfo {year} {2023})}\BibitemShut {NoStop}%
\bibitem [{\citenamefont {Abrams}\ and\ \citenamefont {Lloyd}(1998)}]{24}%
  \BibitemOpen
  \bibfield  {author} {\bibinfo {author} {\bibfnamefont {D.~S.}\ \bibnamefont {Abrams}}\ and\ \bibinfo {author} {\bibfnamefont {S.}~\bibnamefont {Lloyd}},\ }\bibfield  {title} {\bibinfo {title} {Nonlinear quantum mechanics implies polynomial-time solution for np-complete and\# p problems},\ }\href {https://doi.org/10.1103/PhysRevLett.81.3992} {\bibfield  {journal} {\bibinfo  {journal} {Physical Review Letters}\ }\textbf {\bibinfo {volume} {81}},\ \bibinfo {pages} {3992} (\bibinfo {year} {1998})}\BibitemShut {NoStop}%
\bibitem [{\citenamefont {Myrvold}\ \emph {et~al.}(2009)\citenamefont {Myrvold}, \citenamefont {Christian},\ and\ \citenamefont {Hardy}}]{25}%
  \BibitemOpen
  \bibfield  {author} {\bibinfo {author} {\bibfnamefont {W.~C.}\ \bibnamefont {Myrvold}}, \bibinfo {author} {\bibfnamefont {J.}~\bibnamefont {Christian}},\ and\ \bibinfo {author} {\bibfnamefont {L.}~\bibnamefont {Hardy}},\ }\bibfield  {title} {\bibinfo {title} {Quantum gravity computers: On the theory of computation with indefinite causal structure},\ }\href {https://doi.org/10.1007/978-1-4020-9107-0_21} {\bibfield  {journal} {\bibinfo  {journal} {arXiv:Quant-Ph/0701019}\ }\textbf {\bibinfo {volume} {73}},\ \bibinfo {pages} {379} (\bibinfo {year} {2009})}\BibitemShut {NoStop}%
\end{thebibliography}%

\end{document}